\documentclass[useAMS,usenatbib]{mn2e} \input{epsf}

\newif\ifAMStwofonts
\AMStwofontstrue

\usepackage{longtable}
\usepackage{lscape}
\usepackage{float}
\usepackage{url}
\usepackage[english]{babel}
\usepackage[utf8]{inputenc}
\usepackage[varg]{txfonts}
\usepackage{graphicx}
\usepackage{txfonts}
\usepackage{natbib}

\makeatletter
\newlength{\abovecaptionskip}
\makeatother
\usepackage{threeparttable}

\def\lesssim{\mathrel{\hbox{\rlap{\hbox{\lower4pt\hbox{$\sim$}}}\hbox{$<$}}}}
\def\gtrsim{\mathrel{\hbox{\rlap{\hbox{\lower4pt\hbox{$\sim$}}}\hbox{$>$}}}}

\def\l_lsun{$\log{L/\rm L_{\odot}}$~}
\def\masa_msun{$M/ \rm M_{\odot}$~}

\def\m_mstar{$M/M_{*}$~}

\title[Formation of a very massive millisecond pulsar]{Binary evolution leading
to the formation of the very massive neutron star in the J0740+6620 binary
system}

 \author[M. Echeveste, M. L. Novarino, O. G. Benvenuto, \& M. A. De Vito]
 {
 M. Echeveste\thanks{Fellow of the Consejo Nacional de Investigaciones Cient\'{\i}ficas y T\'ecnicas (CONICET)},
 M. L. Novarino$^{\star}$,
 O. G. Benvenuto\thanks{Member of  the Carrera del Investigador
 Cient\'{\i}fico, Comisi\'on de  Investigaciones Cient\'{\i}ficas de la  Provincia
 de Buenos Aires (CIC). Email: obenvenu@fcaglp.unlp.edu.ar},
 M. A. De  Vito\thanks{Member of  the Carrera del Investigador
 Cient\'{\i}fico of CONICET. Email: adevito@fcaglp.unlp.edu.ar},\\
 Instituto de Astrof\'{\i}sica de La Plata, IALP, CCT-CONICET-UNLP, Argentina and\\
 Facultad de Ciencias Astron\'omicas y Geof\'{\i}sicas, Universidad
 Nacional de La Plata (UNLP),\\ Paseo del Bosque S/N, B1900FWA, La Plata,
 Argentina}

\begin{document}

\date{14 May}

\pagerange{\pageref{firstpage}--\pageref{lastpage}} \pubyear{2019}

\maketitle \label{firstpage}

\begin{abstract}
We study the evolution of close binary systems in order to account for the
existence of the recently observed binary system containing the most massive
millisecond pulsar ever detected, PSR J0740+6620, and its ultra-cool helium
white dwarf companion. In order to find a progenitor for this object we compute
the evolution of several binary systems composed by a neutron star and a normal
donor star employing our stellar code. We assume conservative mass transfer. We
also explore the effects of irradiation feedback on the system. We find that
irradiated models also provide adequate models for the millisecond pulsar and
its companion, so both irradiated and non irradiated systems are good
progenitors for PSR J0740+6620. Finally, we obtain a binary system that evolves
and accounts for the observational data of the system composed by PSR J0740+6620
(i.e. orbital period, mass, effective temperature and inferred metallicity of
the companion, and mass of the neutron star) in a time scale smaller than the
age of the Universe. In order to reach an effective temperature as low as
observed, the donor star should have an helium envelope as demanded by
observations.
\end{abstract}


\section{Introduction}

Millisecond pulsars (MSP) are neutron stars (NS) with very short and stable spin
period (P $\leq 30$~ms, $\dot{\textrm{P}}\leq 10^{-19}$). They are widely
considered as an useful tool for testing fundamental physics and studying
gravitational waves emission, binary stellar evolution, and even the properties
of the interstellar medium. At present we know that there are several MSPs in
binary systems \citep{2005AJ....129.1993M}, most of them orbiting together with
another NS or with a white dwarf (WD). If the orbit of the system is nearly edge
on, the mass of the NS and its companion can be inferred with high precision by
radio timing observations, with the measurement of the relativistic Shapiro
delay \citep{1964PhRvL..13..789S}. Knowing the NS mass is critical for
understanding the interior of these stars, since the mass provides a strong
constraint in the equation of state of matter at supra-nuclear densities; see,
e.g., \citet{2004Sci...304..536L}.

Sometime ago it was widely believed that NSs had a typical mass value of
$\approx$1.4~$M_{\odot}$ (usually called the ``canonical mass value''). Nevertheless, recently it has been possible to measure
several MSP's masses largely exceeding this value. Particularly relevants are
the cases of PSR~J1614-2230 with $1.928 \pm 0.017$~$M_{\odot}$
(\citealt{2010Natur.467.1081D}; \citealt{2016ApJ...832..167F}) and
PSR~J0348+0432 with $2.01\pm 0.04$~$M_{\odot}$ \citep{2013Sci...340..448A}. For
an updated review on this topic see  \citet{2019Univ....5..159L}. 

Very recently, \citet{2019NatAs.tmp..439C} announced the observation of
the most massive NS ever detected. It belongs to the PSR~J0740+6620 system,
discovered by  \citet{2014ApJ...791...67S}. This NS has a  mass of
$2.14^{+0.10}_{-0.09}~M_{\odot}$ (68\% credibility interval) and 
$2.14^{+0.20}_{-0.18}~M_{\odot}$ (95.4\% credibility interval), with a spin
period $P_S= 2.89$~ms. The mass of the companion is $M_2= 0.258(8)~M_{\odot}$ and the orbital period of the binary is $4.77$~d.
Independently, \citet{2019MNRAS.485.3715B} reported observations of the
MSP~J0740+6620 companion, finding that it is an ultra-cool WD with a pure helium
atmosphere and an effective temperature $T_{eff} \leq 3500$~K.

PSR~J0740+6620 represents an extremely interesting object of study since it is
likely the most massive NS known to date. One of the most straightforward and
relevant application of the observed mass value is to put a fundamental
constraint on the equation of state (EOS) of dense matter. For a given EOS we
can integrate the equations of General Relativistic stellar structure (known as
the Tolman, Oppenheimer and Volkoff equations) and find its corresponding mass
vs. radius relation. In particular, it is possible to calculate the maximum mass
object it allows to exist. Many of the several proposed EOSs are unable to
support  $2.14^{+0.10}_{-0.09}~M_{\odot}$ against gravitational collapse (see,
for example, Fig.~2 of \citealt{2004Sci...304..536L}). So, the very existence of
a NS as massive as that present in PSR~J0740+6620 is sufficient to discard them
as unrealistic.

On another side, it is interesting to find the way the system PSR~J0740+6620 could have been formed. This is the goal of the present paper. For this purpose, we study the evolution of a set of close binary systems composed by a normal, non degenerate star together with a NS\footnote{As usual, the NS and donor star
are indicated with subscripts 1 and 2 respectively} that accretes mass from its companion. The companion acts as a donor star, and eventually evolves into a WD. 
This is a standard scenario explored by several authors as, e.g.,
\citet{2002ApJ...565.1107P}, \citet{2005MNRAS.362..891B}, and
\citet{2018MNRAS.479..817T}. The NS becomes a MSP as a result of the increase of its mass and angular
momentum due to mass transfer from its companion \citep{1982Natur.300..728A}.
This standard scenario
predicts a long and stable episode of mass transfer as a consequence of the
nuclear evolution of the donor star and angular momentum losses, and a small
number of RLOFs due to thermonuclear flashes \citep{2005MNRAS.362..891B}.

 \citet{2003ApJ...597.1036P} first performed a study that combines binary population synthesis in the Galactic disk and detailed evolutionary calculations of low - and intermediate-mass X-ray binaries (L/IMXBs). In their comprehensive work, the authors computed distributions of the orbital periods, donor masses, mass accretion rates of L/IMXBs, and orbital-period distributions of binary MSPs. In particular, they studied the distribution of NS masses resulting from close binary evolution. Their calculations lead to the formation of NSs as massive as $\sim 2.5~M_{\odot}$.

Employing the MESA code, \citet{2011ApJ...732...70L} computed an extensive grid of binary evolutionary tracks with initial donor masses in the range of $1 - 4~M_{\odot}$ and initial orbital periods between $10$ and $250$~h. Of particular interest is their Figure~4, where it is shown that NS  masses greater than $2~M_{\odot}$ are possible for systems with the low mass companions ($\sim 0.15 - 0.25~M_{\odot}$), and orbital periods between $10$ and $80$~h. They applied their results to PSR~J1614-2230, which has a $1.97 \pm 0.04~M_{\odot}$ NS \citep{2010Natur.467.1081D}, an orbital period of $8.7$~days, and a companion star of $0.5~M_{\odot}$. The authors claim that an initial $1.4~M_{\odot}$ NS together with a $3.4 - 3.8~M_{\odot}$  donor star mass evolves and reproduces the present configuration of PSR~J1614-2230. However the final NS high mass value is not easily reached. Indeed, to fit the current mass of the NS, it must have initially been at least of $1.6 \pm 0.1~M_{\odot}$. In the same way, the calculations made by \citet{2011MNRAS.416.2130T}
require a NS that was born with a mass greater than the canonical value of $1.4~M_{\odot}$ to fit the observed pulsar mass in PSR~J1614-2230.

The standard model includes neither evaporation of the donor star by radio
pulsar irradiation nor X-ray irradiation feedback. Evaporation leads the donor
star to enhanced mass losses. 
 This phenomenon was studied by \citet{1989ApJ...336..507R} and  \citet{1989ApJ...343..292R}.
In these papers, the authors studied the effect of evaporation wind in the case of binary systems composed by a NS with very light companions ($< 0.1~M_{\odot}$) due to various types of radiation, during accretion phase or when this process has ended. One of the main objectives of their research was to find a possible explanation for the existence of isolated millisecond pulsars without a visible companion that has acted as its donor in a recycled scenario. The role of evaporation is essential to depict the evolution of certain binaries to the black widow state (see, e.g., \citealt{2012ApJ...753L..33B}). 

On the other hand, irradiation feedback occurs
during RLOF episodes, when matter falls onto the NS and releases X-ray
irradiation that illuminates the donor star. \citet{1991Natur.350..136P} first studied the effect of irradiation in LMXBs. For stars that have a thick enough outer convective zone, this phenomenon makes their structure to change
considerably making its effective surface to become smaller. Accordingly, in
some cases the donor star is unable to sustain the RLOF and becomes detached.
Subsequent nuclear evolution may lead the donor star to experience RLOF again,
undergoing a quasi-cyclic behaviour (\citealt{1993A&A...277...81H}; \citealt{2004A&A...423..281B}; \citealt{2014ApJ...786L...7B}).  This process may affect the evolution of the LMXBs, explaining the classical discrepancy between the millisecond pulsar and LMXB lifetimes \citep{2003ApJ...597.1036P}. 

Here, we shall show that it is possible to account for the masses, the orbital period, and the characteristics of the WD of PSR~J0740+6620 system, provided that the system has a very low metallicity. Besides, we show that the NS in this system can reach its large mass without the necessity of being initially more massive than the canonical value.
We shall also consider the capability of binary evolution models considering
irradiation feedback to provide a plausible scenario for the formation of the PSR~J0740+6620 system. 

The remainder of this paper is organised as follows. In
Section~\ref{sec:codigo_numerico}, we describe our stellar
code. In Section~\ref{sec:calcu}, we expose the results we obtained and present a progenitor for
PSR~J0740+6620. Lastly, in Section~\ref{sec:conclu}, we review the main results presented in this paper and we give some concluding remarks.

\section{Our numerical model} \label{sec:codigo_numerico}

Our research was performed using the binary evolutionary code presented in
\citet{2003MNRAS.342...50B}. This code was updated by the inclusion of
evaporation of the donor star and irradiation feedback. Every time the system is
in a Roche Lobe OverFlow (RLOF) state, the code works in a fully implicit way,
solving the donor star's structure together with the mass transfer rate, the
increase/decrease of the mass of both stars in the system and the evolution of
the orbital semi-axis. This method is numerically stable and allows for the
calculation of mass transfer cycles \citep{2012ApJ...753L..33B}. When the system
is detached, the code employs the standard Henyey technique. For further
details, we refer the reader to \citet{2014ApJ...786L...7B} and references
therein. In our studies the normal star is the donor, i.e., the star that suffers  RLOF, and the NS acts as an accreting compact object. In this work we shall take irradiation feedback into account but ignore evaporation since at these stages it is expected to be
not relevant. 

\section{Numerical Results} \label{sec:calcu}

In order to find a plausible progenitor for the PSR~J0740+6620 system, we
computed the evolution of binary systems initially composed by a NS and a normal star on the Zero Age Main Sequence. We considered donor stars with initial mass $M_2= 1\ M_{\odot}$, hydrogen mass abundance of $X= 0.7381$ and metallicities of $Z= 0.0174$ (the Solar value, see \citealt{2009ARA&A..47..481A}), 0.0010, 0.0003 and 0.0001. The initial NS mass is assumed to be $1.4 M_{\odot}$ in all our simulations.

\begin{figure}
    \centering
    \includegraphics[width=0.5\textwidth]{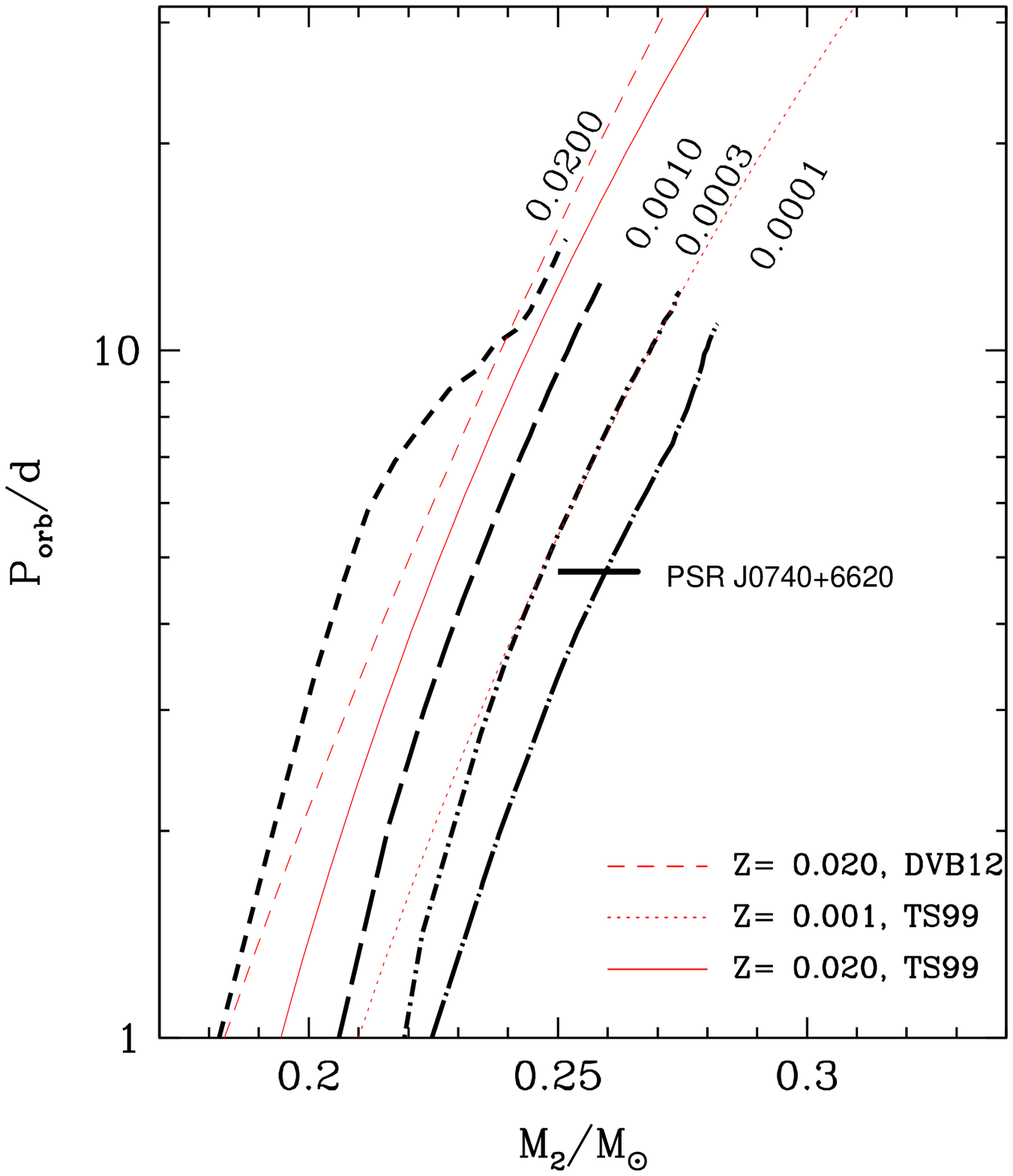}
    \caption{Final orbital period of the system as a function of the final mass of the donor star for different metallicities. The metallicity of the donor star is indicated for each curve. Black curves were calculated for this work and the red dot curve labeled as DVB12 corresponds to \citet{2012MNRAS.421.2206D}, whereas the other red curves were taken from \citet{1999A&A...350..928T}. The horizontal thick line labeled as PSR~J0740+6620 represents the observational data of this system. }
    \label{fig:periodos}
\end{figure}

Initially, in constructing these models we considered different values for the fraction $\beta$ of the matter transferred from the donor that is eventually accreted by the NS. The accretion onto the NS is limited by the so called Eddington accretion rate $\dot{M}_{Edd}= 2 \times 10^{-8}\ M_{\odot} y^{-1}$ \citep{2002ApJ...565.1107P}. The choice of the parameter $\beta$ has a direct impact on the final NS mass, but has a minor effect on the evolution of the donor star. After some exploration of the parameters space we decided to assume conservative mass transfer, i.e. $\beta= 1$ in all our final calculations. Of course, lower values of $\beta$ and higher initial masses $M_{2}$ are possible and would eventually give other different solutions for the progenitor of PSR~J0740+6620. Nevertheless, it is not the aim of this paper to provide a whole family of plausible solutions for the problem at hand, but to demonstrate that at least one does exist. If $\beta<1$ we need to consider initially more massive donor stars because they undergo higher mass transfer rates on shorter RLOF episodes. Thus, the threshold imposed by $\dot{M}_{Edd}$ will have a larger impact, since a fraction of the transferred mass is lost away from the system and the
condition that the NS has to grow up to the observed mass in PSR~J0740+6620
system is more difficult reach.

\begin{figure}
    \centering
    \includegraphics[width=0.5\textwidth]{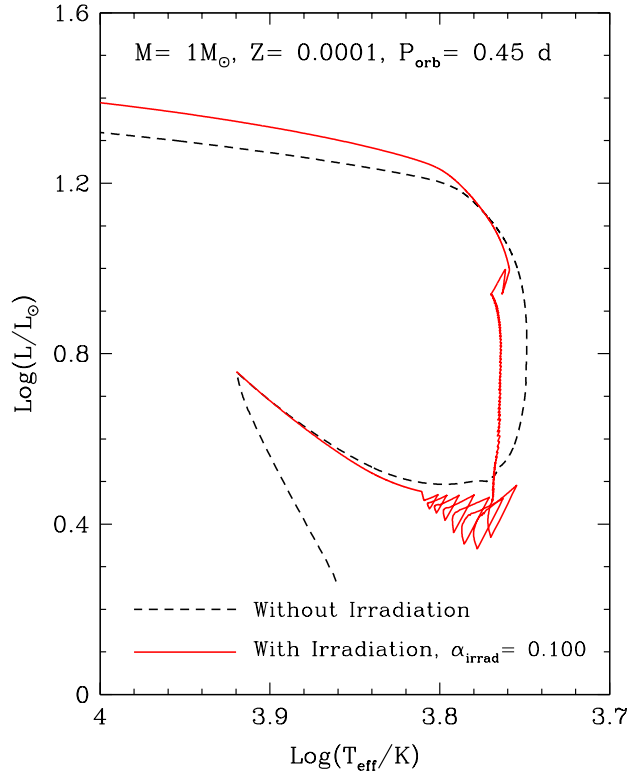}
    \caption{HR diagram for the donor star with initial mass $M_2$= 1~$M_{\odot}$, Z= 0.00010 and initial orbital period $P_{orb}$= 0.45~days. The continuous red line corresponds to the irradiated model with $\alpha_{irrad}= 0.10$ whereas the black dot dashed curve corresponds to non irradiated one. }
    \label{fig:hr_irrad}
\end{figure}

It is well known that for systems that have donor stars with masses not too low, there is a rather well defined relation between the final donor mass and the orbital period, $M_2$--$P_{orb}$ (see, e.g., \citealt{1995MNRAS.273..731R}) We employed these relations to find a system
able to account for the observed masses and orbital period of PSR~J0740+6620.
Fig.~\ref{fig:periodos} shows these relations for the metallicity values
computed for this work. Each point on these curves represents the final state of
the evolutionary track of one binary system. The characteristics observed for
PSR~J0740+6620 are reached by a system with Z= 0.00010. This is in qualitative
concordance with the suggestion made by \citet{2019NatAs.tmp..439C} based on the
relations presented by \citet{1999A&A...350..928T}. Following our calculations,
we conclude that a system that undergoes conservative mass transfer and
initially has $M_2= 1\ M_{\odot}$, $M_{NS}=1.4\ M_{\odot}$, Z= 0.00010 and
$P_{orb}$= 0.45~days represents a plausible progenitor for PSR~J0740+6620.

\begin{figure}
    \centering
    \includegraphics[width=0.5\textwidth]{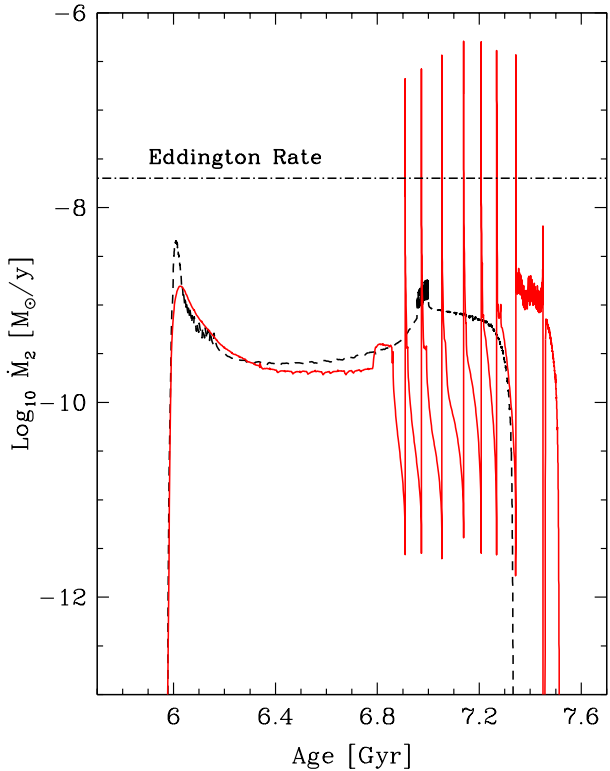}
    \caption{Donor mass transfer rate as a function of time for a system with initial mass $M_2$= 1~$M_{\odot}$ and initial orbital period $P_{orb}$= 0.45~days. The continuous red line corresponds to a system with irradiation feedback with $\alpha_{irrad}= 0.10$ and the black dashed curve corresponds to a system without irradiation. The horizontal line indicates the Eddington Rate  }
    \label{fig:Mloss_irrad}
\end{figure}

\begin{figure}
    \centering
    \includegraphics[width=0.5\textwidth]{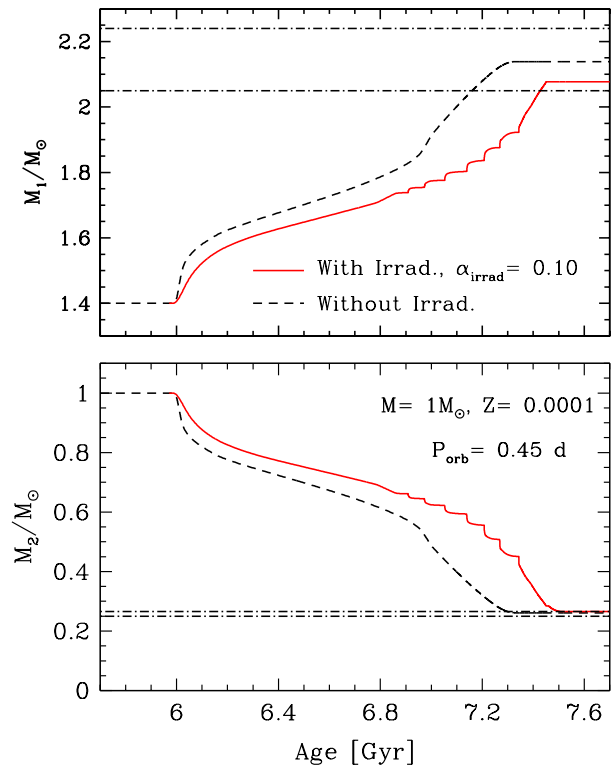}
    \caption{The masses of the components a function of time. The continuous red line corresponds to the irradiated system with $\alpha_{irrad}= 0.10$ and the black dashed curve corresponds to the non irradiated system. The top (bottom) panel shows the evolution of the mass of the NS (donor star). The horizontal lines represent the limits of the observed mass respectively. }
    \label{fig:Masas_irrad}
\end{figure}

In addition, as stated above (\S~\ref{sec:codigo_numerico}), we took into
account the effects of irradiation feedback on the evolution of the system
described above (with initial $M_2$= 1~$M_{\odot}$, initial $P_{orb}$= 0.45~days
and Z= 0.00010). We considered the case of $\alpha_{irrad}= 0.10$ which represents an intermediate case for the effects induced by irradiation on the evolution. Fig.~\ref{fig:hr_irrad} shows the evolution of the system considering irradiation and ignoring it. The irradiated system undergoes some mass transfer cycles. As it can be seen in Fig.~\ref{fig:Mloss_irrad}, during the cyclic stage of mass transfer of irradiated models, the mass transfer rate from the donor star largely exceeds the Eddington Rate on very short timescales. Thus, even having assumed $\beta=1$, some material is lost from the system. This represents an obvious difficulty for getting high mass values for the accreting NS. A detailed calculation shows that even in this situation, the masses of the components $M_2$ and $M_1$ (NS mass) of the system are still compatible with the observations made by \citet{2019NatAs.tmp..439C} (see Fig.~\ref{fig:Masas_irrad}). Hence, irradiation is not an unavoidable ingredient to successfully reach the observed parameters (masses and orbital period). So, for studying the present evolutionary state of PSR~J0740+6620 this phenomenon may be neglected. 

In what follows, we shall analyse three different conditions for the evolution
of the donor star ignoring irradiation feedback\footnote{It is a fortunate
situation that this is adequate for our purposes, since irradiated models are
{\it very} time consuming.}. After the donor star detaches from its Roche Lobe, it has a thick hydrogen-rich envelope while the interior has a helium
composition. Thus, initially we analysed the cases of donor models considering
and ignoring diffusion. As we shall show below, these models do not cool down
fast enough to account for the observations of a very cool helium-dominated
atmosphere WD made by \citet{2019MNRAS.485.3715B}. To reach such a state we had
to  remove all the remaining H in the donor star envelope. We did it when the
donor attains its maximum luminosity on the other tracks assuming it has been
converted into helium.

\begin{figure*}
    \centering
    \includegraphics[angle=-90,width=0.80\textwidth]{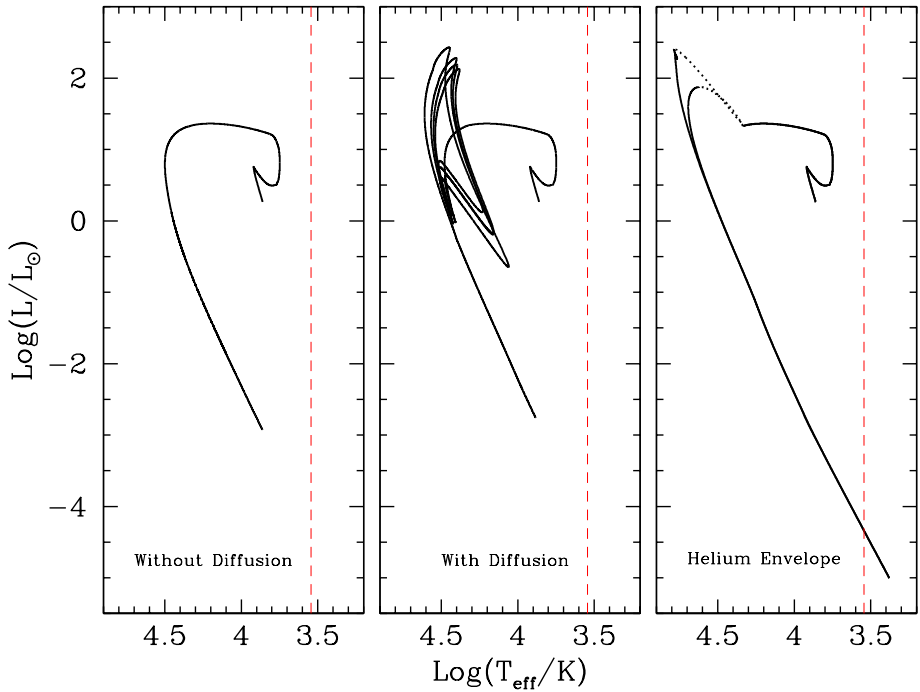}
    \caption{Hertzsprung-Russell diagram for the evolution of a donor star with initial mass $M_2$= 1~$M_{\odot}$ and initial orbital period $P_{orb}$= 0.45~days. The left panel shows the evolution of the system without diffusion, the middle panel corresponds to the case with diffusion and the right panel shows the system which has lost all its H. In the latter, the two tracks correspond to the details of the artificial procedure performed to remove the remaining H
    (the fewer the number of models we employ to switch from H to He, the higher the maximum luminosity). Of key relevance is that these details have no impact on the subsequent evolution of the object.} The temperature observed for PSR~J0740+6620, denoted with a vertical red dot line. 
    \label{fig:hr}
\end{figure*}

\begin{figure}
    \centering
    \includegraphics[width=0.5\textwidth]{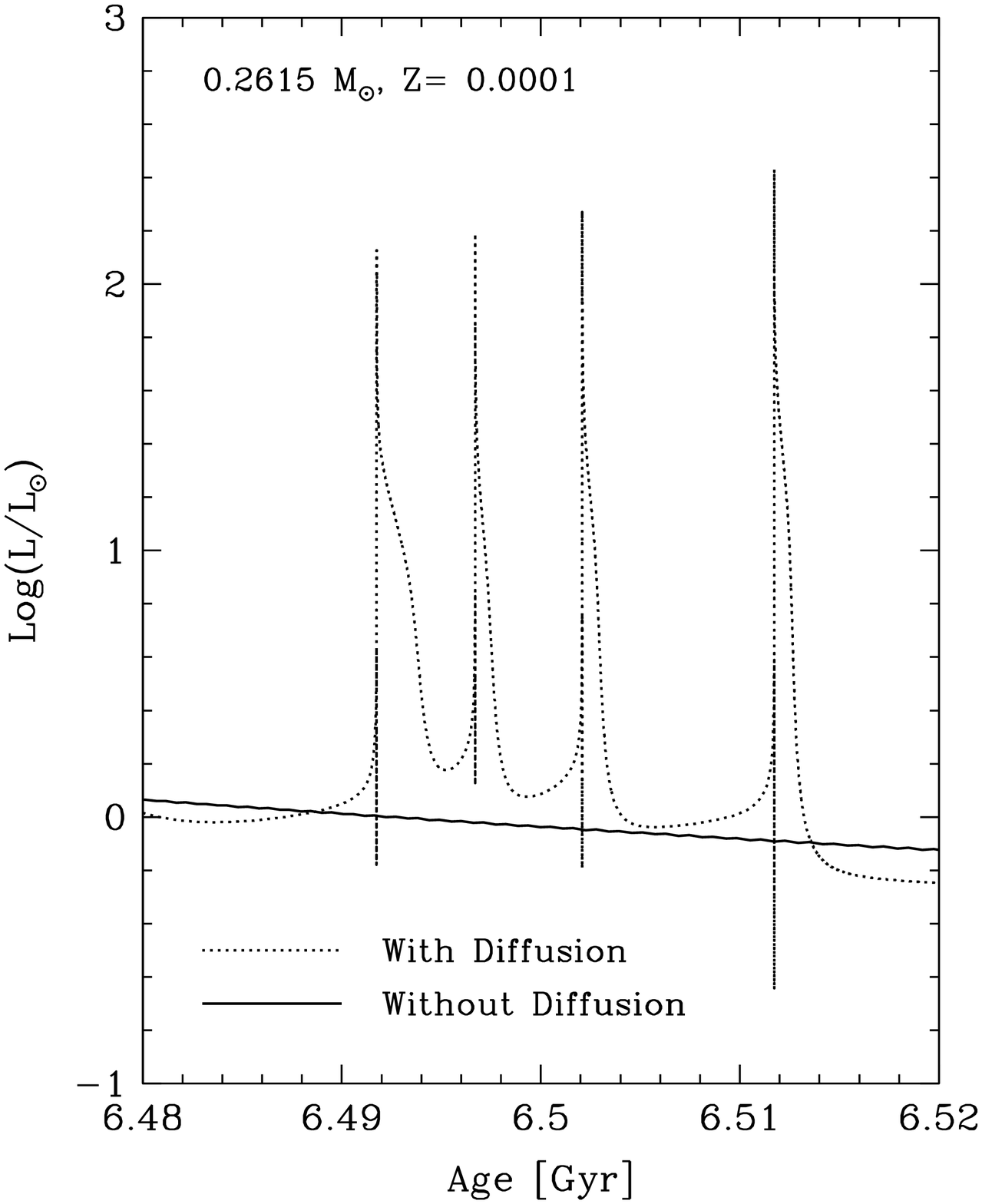}
    \caption{The luminosity as a function of time for the evolution of the donor star during the hydrogen shell thermonuclear flashes caused by the diffusion processes. }
    \label{fig:flashes}
\end{figure}

The evolutionary tracks for these three cases are presented in
Fig.~\ref{fig:hr}. The calculations without diffusion show a fast evolution to
high effective temperature values and a subsequent slow and smooth cooling as a
WD. Calculations were stopped at ages in excess of the Hubble time. Even so, the
donor star remained far hotter than observed, making it to be incompatible with
observations. It can be expected that models considering diffusion are more
promising. Diffusion provides an hydrogen tail that reaches very hot layers
inducing the occurrence of several thermonuclear flashes at the bottom of that
rather outer layers (se, e.g., \citealt{2004MNRAS.352..249B}). Indeed, much
hydrogen is burnt out making the star to cool down faster than the model without
diffusion (see Fig.~\ref{fig:flashes}). However, as in the former case, models
considering diffusion still remain hotter than observations for the age of the
Universe. Notice that, apart from the temperature and luminosity, the outer
stellar layers are still hydrogen rich.

\begin{figure}
    \centering
    \includegraphics[angle=270,width=0.5\textwidth]{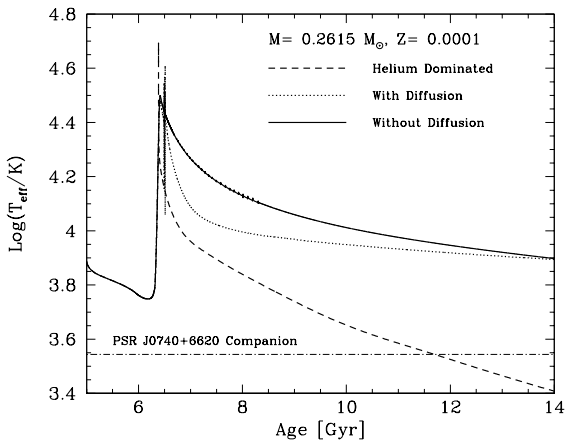}
    \caption{Effective Temperature of the donor star as a function of time for the three different cases  (H-rich envelope without and with diffusion, and He-rich envelope) shown in Fig.~\ref{fig:hr}. The temperature observed for PSR~J0740+6620 companion is denoted with an horizontal line.}
    \label{fig:Teff-t}
\end{figure}

\begin{figure}
    \centering
    \includegraphics[angle=270,width=0.5\textwidth]{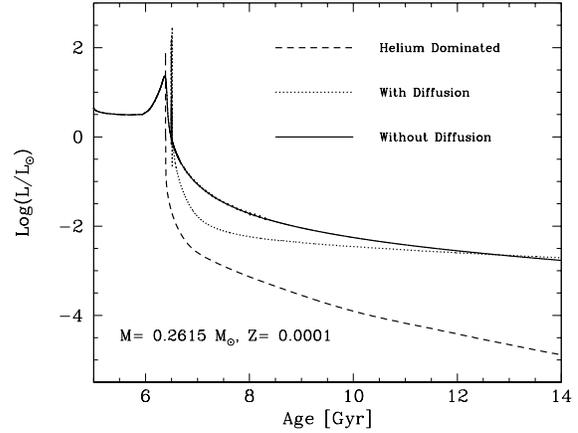}
    \caption{Luminosity of the donor star as a function of time for the three different cases  (H-rich envelope without and with diffusion, and He-rich envelope) presented in Fig.~\ref{fig:hr}.}
    \label{fig:lumi}
\end{figure}

The donor star model in which we have removed all H at its envelope initially
reaches a very high temperature and luminosity because its outer layers became
suddenly much more transparent. This should be considered as an unrealistic
consequence of the procedure.  This very high luminosity depends on the details of the artificial procedure performed to remove the remaining H. Fortunately, these details have no impact on the subsequent evolution of the object. The model quickly reaches a cooling track for an helium-rich WD finally reaches a temperature compatible with the
observations made by \citet{2019MNRAS.485.3715B} in a time scale smaller than
the age of the Universe. On the contrary, donor stars which still have H in
their envelopes do not cool enough to become compatible with observations.
Fig.~\ref{fig:Teff-t} shows the evolution of the temperature in the three cases
(H-rich envelope without and with diffusion, and He-rich envelope). After the
age of 6.5~Gyrs, the curves begin to separate and the system which has a He
dominated envelope cools down faster than the other models. Something similar
happens with the luminosity (Fig.~\ref{fig:lumi}). For an He dominated envelope,
luminosity suffers a pronounced decay, caused by the lack of H which acts as an
insulator. 

It can be considered that the so called ``born-again'' scenario proposed long
time ago by \citet{1983ApJ...264..605I} is a promising process to make the star
to loss all its hydrogen by a nuclear burning process. Although we found several
hydrogen burning flashes, they were not strong enough to burn out all the
remaining hydrogen. In any case, for PSR~J0740+6620 some process should have
happened to remove the remaining hydrogen, but the detailed physical process is
still not clear at this moment.

\section{Discussion and Conclusions} \label{sec:conclu}

In this paper our motivation was to explore a possible origin for the very
massive NS in PSR~J0740+6620. Using the binary evolutionary code developed by
our group, we searched for plausible progenitors for the binary system that
contains this MSP. We considered a conservative system, i.e. all the matter lost
by the donor, when it is below the $M_{Edd}$ rate, is accreted by the NS
($\beta=1$), and examined different initial  values for the mass of the donor
star, the orbital period and  metallicites. We analysed the effects of
irradiation feedback on the evolution of the system, finding that both
irradiated and non irradiated models provide plausible progenitors for
PSR~J0740+6620.  However, binary evolution predicts a hydrogen rich envelope.
Thus, we explored diffusion effects on the donor star and found that stars which
have H-rich envelopes do not cool down fast enough to reach the very low
effective temperature  observed for its companion \citep{2019MNRAS.485.3715B}.
If we remove the remaining hydrogen of the envelope of the donor star, we
verified that the compact remnant of the donor star cools down fast enough to
reach the observed effective temperature in a time scale smaller than the age of
the Universe. 

We found that the model with initial masses $M_1= 1.4~M_{\odot}$ and $M_2$=
1~$M_{\odot}$, orbital period $P_{orb}$=0.45~days, and Z= 0.00010, accounts for
the evolutionary status of the system observed by \citet{2019NatAs.tmp..439C}. Also, from an evolutionary point of view, we found that the envelope of the white
dwarf should be hydrogen free, which is in nice agreement with the composition
observed by \citep{2019MNRAS.485.3715B}. 


In our calculations we have found that the donor star experiences few thermonuclear flashes during which lead to a sudden increase of luminosity. These flashes do not burn all its remaining hydrogen leading to large the discrepancy with observations.  Evidently, in our calculations some physical ingredient has a largely underrated relevance, or is simply lacking. One possibility is that the effects of diffusion have been underestimated and in reality they lead to stronger thermonuclear flashes. If so, much of the hydrogen would be burnt out but also some of it may be lost in a later flash driven RLOF. Another possibility is that evaporation of the donor is indeed non-negligible as we have assumed in calculations. All these possibilities warrant a future detailed exploration.

We want to thank our anonymous referee for his/her corrections and suggestions that have helped us improve the original version of this paper.

\label{lastpage}

\end{document}